\documentclass[useAMS,usenatbib,psfig]{mn2e}
\input{epsf}
\input psfig.sty

\newcommand{\aap}    {A\&A}
\newcommand{\apjs}   {ApJS}
\newcommand{\apj}    {ApJ}
\newcommand{\apjl}   {ApJL}
\newcommand{\aj}     {AJ}
\newcommand{\mnras}  {MNRAS}

\title[Alignments of Galaxy Group Shapes with LSS]{Alignments of Galaxy Group Shapes with  Large Scale Structure}
\author[D. J. Paz et al.]
{Dante J. Paz$^1$\thanks{E-mail:dpaz@oac.uncor.edu}, 
 Mario A. Sgr\'o$^1$ ,
 Manuel Merch\'an$^1$,
 Nelson Padilla$^2$\\
 $^1$ Instituto de Astronom\'{\i}a Te\'orica y Experimental (UNC-CONICET), Observatorio Astron\'omico de C\'ordoba, \\
 Laprida 854, C\'ordoba, X500BGR, Argentina\\
 $^2$ Departamento de Astronom\'\i a y Astrof\'\i sica, Universidad Cat\'olica
 de Chile, Vicu\~na Mackenna 4860, Santiago, Chile.\\
}

\begin{document}
\date{Accepted ---- . Received --- }

\pagerange{\pageref{firstpage}--\pageref{lastpage}} \pubyear{2010}

\maketitle

\label{firstpage}

\begin{abstract}
   In this paper we analyse the alignment of galaxy groups with the surrounding
   large scale structure traced by spectroscopic galaxies from the Sloan
   Digital Sky Survey Data Release 7. We characterise these alignments by means
   of an extension of the classical two-point cross-correlation function,
   developed by \citet{PaperL}. We find a strong alignment signal between the
   projected major axis of group shapes and the surrounding galaxy
   distribution up to scales of $30$ Mpc h$^{-1}$. This observed anisotropy
   signal becomes larger as the galaxy group mass increases, in excellent
   agreement with the corresponding predicted alignment obtained from mock
   catalogues and $\Lambda$CDM cosmological simulations. These measurements
   provide new direct evidence of the adequacy of the gravitational instability
   picture to describe the large-scale structure formation of our Universe.
\end{abstract}

\begin{keywords}
galaxies: groups: general, dark matter, large-scale structure of universe.
\end{keywords}

\section{Introduction}

\begin{table}
\end{table}

The majority of studies of clusters of galaxies have assumed they are the
visible component of a gravitationally bound system dominated by a dark matter
halo. It is well known that these objects exhibit mildly aspherical shapes,
with a slight preference toward prolate forms. Moreover, it has been shown that
their orientations are related to the surrounding structures such as filaments
and large-scale walls \citep{2005Colberg_Shape, 2005Kasun_Shape,
2006Basilakos_Shape, 2006Allgood_Shape, 2006Altay_Shape, 2007Aragon_Spin,
2007Brunino_Shape, 2007Bett_Shape, 2009Zhang_Spin}.  Results from numerical
simulations by \citeauthor{1993vanHaarlem_Fields}
(\citeyear{1993vanHaarlem_Fields}, see also
\citeauthor{1997Splinter_Ellipticity} \citeyear{1997Splinter_Ellipticity}) have
shown that the origin for such alignments comes from the rearrangement of dark
matter halo axes in the direction from where matter was predominantly accreted
(i.e. last major merger event).  Using numerical simulations, several authors
have found that dark matter halos tend to be more prolate and aspherical when
larger halo masses are considered \citep[][and references
therein]{2005Kasun_Shape,PaperShapes}. This trend is consistent with the
picture described above, and could be easily understood as the result of the
accretion process driven by mergers.  After a merger occurs, dynamical
relaxation will tend to drive a halo closer to isotropy.  Therefore, it is
expected that dynamically younger systems will be more strongly ellipsoidal
than dynamically older ones. Given that the dynamical age of a system depends
on how much time has elapsed from the moment of the last halo merger, it is
expected that high-mass halos will tend to be more elongated than lower mass
halos.\footnote{ Small halos may start their development later than higher mass
systems (\citealt{LPC09}), but the former will still be subject to a lower
number of major mergers (\citealt{DL06}).} \citet{PaperShapes} also found that
this predicted trend is consistent with observational results obtained from
samples of galaxy groups. More recently, \citet{RobotGroups} confirms this
agreement between galaxy group shapes and simulated dark matter halos. Moreover
\citet{RagoneForm} found that less concentrated groups, corresponding to later
formation times, show lower velocity dispersions and higher elongations than
groups of the same mass with higher values of concentration, which
correspond to earlier formation times.  Therefore, in a statistical sense, the
shapes of galaxy groups and their alignments could encode information about the
formation of large-scale structure (LSS).

\citet{Binggeli82} was the first to investigate alignments of pairs of clusters
close to each other.  Taking $44$ Abell groups, Binggeli found that galaxies
separated by up to $30$ Mpc show a strong alignment, while the orientation of a
given cluster is related to the spatial distribution of neighbour systems.
With some exceptions, for example \citet{Struble85} or \citet{Rhee87}, most
papers in the literature confirm this result \citep{Theije95, WestA, Rhee92,
Onuora2000, West95}.  Other authors \citep{Rhee87, WestB, WestA, Richstone92,
Plionis94} also reported both, alignment between neighbour clusters, as well as
alignments with other groups embedded in the same supercluster. For instance,
using $48$ superclusters \citet{WestA} found a tendency for groups to be
aligned on scales as large as $60$ Mpc h$^{-1}$. More recently
\citet{2010Godlowski} analysed the orientation of galaxy groups in the Local
Supercluster, finding strong correlations with the distribution of neighbouring
groups up to scales of about $20$ Mpc.  Using the fourth Release of the Sloan
Digital Sky Survey \citep[SDSS,][]{YorkSDSS}, \citet{2009Wang} report several
types of alignment signals between pairs of neighbouring galaxy groups and
between groups and their surrounding galaxies.

On the other hand, there is an extensive amount of work studying the presence
of alignments between halo shapes and the large scale structure using numerical
simulations \citep{2007Aragon_Spin,Hahn1,2007Brunino_Shape,Cuesta1, Patiri1,
2006Basilakos_Shape}.  Most of these studies perform statistics on the
inclinations of halo axes with respect to directions defined by the surrounding
structure.  This structure is characterised by the particle distribution
through various topological signatures such as filaments, walls or voids.
While these studies agree on the presence of alignments within the $\Lambda$CDM
context, the magnitude of the alignment effect depends strongly on the
definition of the surrounding structure.

In this paper we report measurements of alignments between galaxy group shapes
and the surrounding galaxy distribution in both, real galaxy groups obtained
from the seventh data release of the Sloan Digital Sky Survey (SDSS-DR7,
\citealt{Abazajian_DR7}) and from a numerical simulation. This is carried out
following a methodology similar to those developed in \citet{PaperL}, where the
two-point correlation function was used to quantify the alignment between
galaxy angular momenta and the large scale structure in both, numerical
simulations and real data. In the present work we perform similar measurements,
by using the ``anisotropic'' cross-correlation functions between galaxy groups
and individual galaxies, in the directions parallel and perpendicular to the
major axis of the group shape. Due to we are only interested in the alignment
between galaxy groups and the surrounding large scale structure, we focus our
analysis on observed anisotropies over the two halo term of the two-point
cross-correlation function. The advantage of this approach is the independence
on any particular characterisation of the surrounding structure.  Furthermore,
this modified two-point cross-correlation function allows a robust comparison
between numerical simulations and observational data.  A similar procedure
to the one presented by \citet{PaperL}, was also applied by
\citet{2009Faltenbacher_corr}.

\section[]{$\Lambda$CDM predictions for halo shape-LSS alignments}

In this section we use a numerical simulation to study the anisotropy of the
halo-particle cross-correlation function. The anisotropy signal around halos is
characterised by taking into account the relative orientation of a particle
position with respect to the shape of the central halo. In order to mimic the
observational effects on the anisotropy signal, we also analyse the
distribution of galaxies around galaxy groups on a SDSS mock catalogue by means
of the projected group-galaxy cross-correlation function.  

\subsection{Numerical simulation}

\begin{figure*} 
\begin{picture}(430,250)
\put(-40,0){\psfig{file=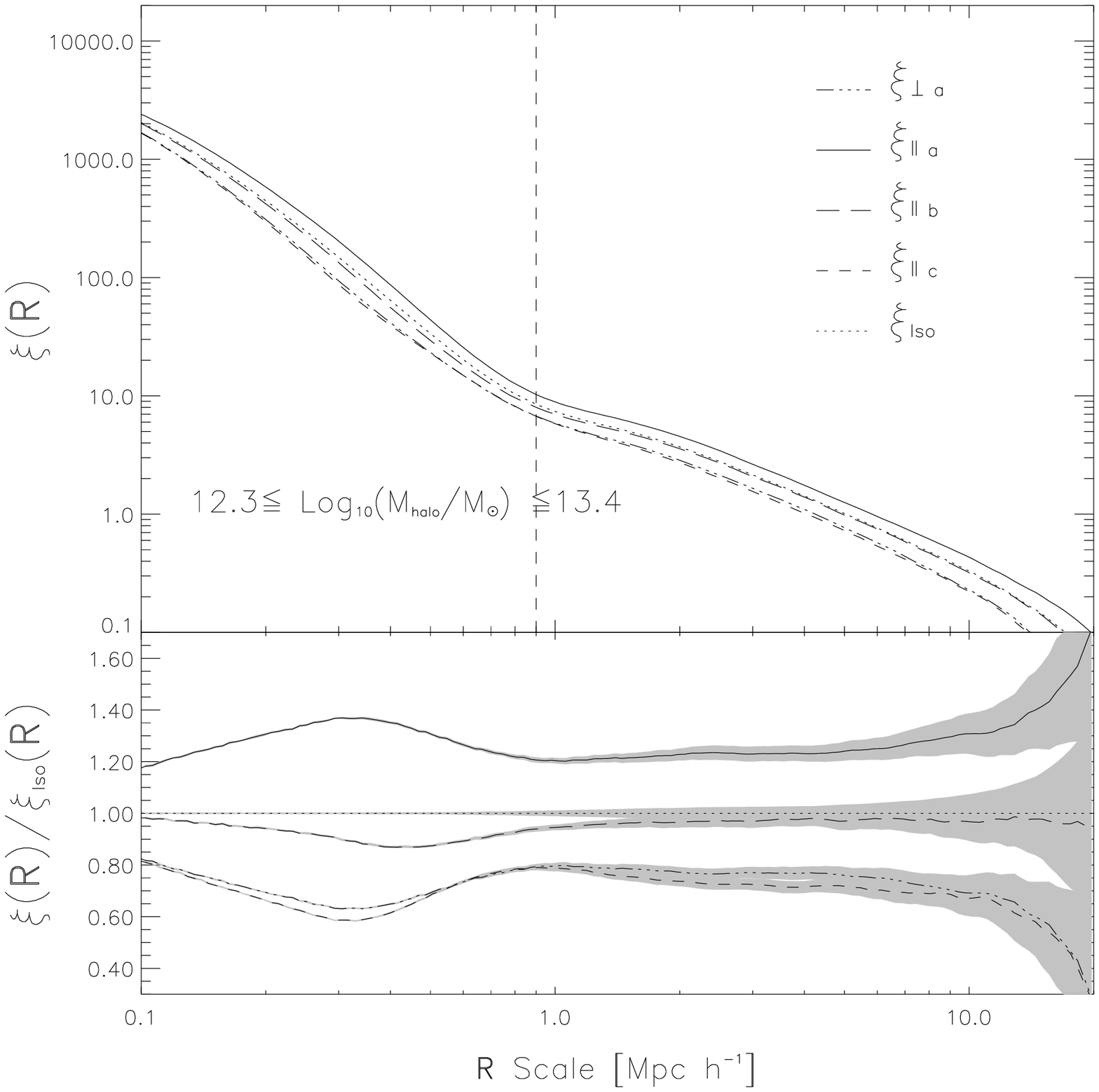,width=9.cm}}
\put(210,5.5){\psfig{file=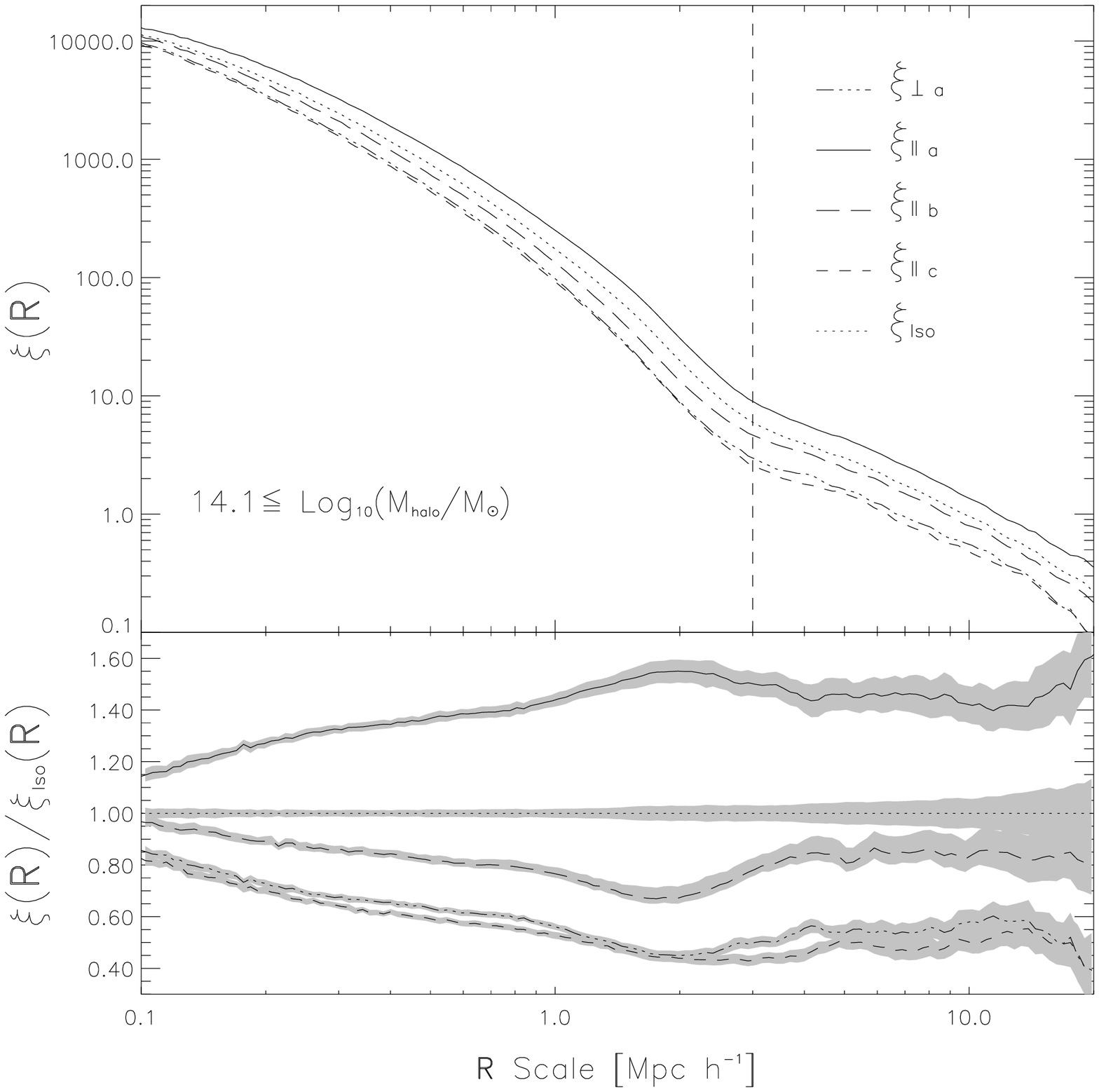,width=8.18cm}}
\end{picture}
\caption{
       Left panel: Spatial Halo$-$Dark-matter correlation function of low mass
       halos in the numerical simulation.  The solid, long dashed and short
       dashed lines show the correlation functions between halos and neighbour
       particles along the directions of the major, intermediate, and minor
       axes of the halo shape ellipsoid, respectively.  The dotted line shows
       the results for the isotropic correlation obtained by using all the
       tracers regardless of the direction (i.e. the classical halo-particle
       cross-correlation function). The dotted-dashed lines correspond to the
       correlation function between halos and neighbour particles in the
       direction perpendicular to the halo shape major axis.  The dashed
       vertical line represents the transition scale between the $1-$ and
       $2-$halo terms of the cross-correlation function.  Shaded areas indicate
       the 1--$\sigma$ standard deviation around each measurement. The lower
       left subpanel shows ratios between the directional and isotropic
       correlation functions shown in the top-left panel.  Right panel: same as
       left panel, for a high mass DM halo subsample obtained from the
       simulation.
}
\label{fig:3D}
\end{figure*}

Throughout this work we use a collisionless numerical simulation covering a
periodic volume of $500^3$ $(h^{-1} {\rm Mpc})^3$. The initial conditions at
redshift $\sim 50$ were calculated assuming a spatially flat low-density
Universe, with a matter and vacuum density parameters $\Omega_{\rm
m}=1-\Omega_{\Lambda}=0.258$, Hubble constant $H_{\circ}=71.9$ km s$^{-1}$
Mpc$^{-1}$, and normalisation parameter $\sigma_{8}=0.796$.  The resulting
particle resolution is $m_{\rm p} =  6.67\times 10^{10}\,h^{-1}\,M_{\odot}$.
The run was performed using the second version of the GADGET code developed by
\citet{Gadget2}, with a gravitational softening of $0.03 h^{-1} {\rm Mpc}$
chosen by following \citet{PowerC}.

The identification of particle clumps was carried out by means of a standard
friends-of-friends algorithm with a percolation length given by $l=0.17$
$\bar{\nu}^{-1/3}$, where $\bar{\nu}$ is the mean number density of DM
particles. In order to obtain reliable measurements of halo shapes (as is
described below), we only consider halos with at least $30$ particle members.
Consequently the resulting halo sample has a minimum mass threshold of
$\sim2\times10^{12}\,h^{-1}\,M_{\odot}$. Nevertheless we are mainly interested
in the observed galaxy group mass range, that is between $10^{13}$ to
$10^{15}\,h^{-1}\,M_{\odot}$ (see figure \ref{fig:mass}). Therefore our results
can be associated with masses ranging from small galaxy groups to galaxy
clusters.

Several authors \citep[][and references therein]{1992Warren, 1998Thomas,
2005Hopkins, 2006Allgood_Shape, 2005Kasun_Shape, LauShapes, PaperShapes} have
analysed the properties of halo shapes using the best-fitting ellipsoid to the
spatial distribution of halo members.  Following this standard method, we
calculate the shape tensor for each dark-matter halo using the positions of
each of its particle members. This can be written as a symmetric matrix,
\begin{equation}
I_{ij}= (1/N_h)\sum_{\alpha=1}^{N_h} X_{\alpha i} X_{\alpha j},
\label{eq:shape} \end{equation}
where $X_{\alpha i}$ is the $i^{th}$ component of the displacement vector of a
particle $\alpha$ relative to the centre of mass, and $N_{h}$ is the number of
particles in the halo. The matrix eigenvalues correspond to the square of the
axis ($a$, $b$, $c$ where $a>b>c$) of the characteristic ellipsoid that
best describes the spatial distribution of the halo members. Our alignment
analysis is based on the eigenvector directions ($\hat{a}$, $\hat{b}$ and
$\hat{c}$).  A fixed $b/a=1$ with an arbitrary value of $c/b$, corresponds to
perfect oblate ellipsoids. On the other hand, systems with fixed $c/b=1$ are
perfect prolate ellipsoids. A system with $b/a<c/b$ is associated to a general
triaxial ellipsoid with prolate tendency, while the opposite case, $b/a>c/b$,
corresponds to a predominantly oblate ellipsoid.

Since we will apply a statistical method to observations, we also construct a
synthetic SDSS catalogue in order to test the effects introduced by
observational biases. By applying the SDSS selection function on our simulation
box we produce a mock catalogue \citep[see for instance,][]{Colemock}.  For
simplicity, and because our goal is merely to match the geometry and depth of
the catalogue, we compute the radial selection function of simulation particles
by randomly assigning them an absolute magnitude.  We use different
Schechter luminosity functions to apply SDSS r-band magnitudes to the
particles. For particles outside groups we use the parameters $\phi_*=0.0149 \;
M_*=-20.44\; \alpha=-1.05$ \citep{BLum}, whereas for particles inside groups,
we use different Schechter parameters, which depend on the halo mass following
\citep{2006ZMM}. The flux-limit selection of the mock catalogue is then
obtained by imposing the same upper apparent magnitude threshold which affects
the observational sample (in our case we consider galaxies with $m < 17.5$, in
both real and mock catalogues). In order to determine the angular selection
function of the survey, we divided the sphere in $\sim 7,700,000$ pixels using
the SDSSPix software. This pixelisation scheme was specifically designed for
the SDSS geometry \citep[see for instance][]{SDSSpix}. We then compute an
initial binary mask by setting to $1$ all pixels with at least one galaxy
inside, and $0$ the remaining pixels. Given that the pixel size is quite
smaller than the mean inter-galaxy separation, we smooth the initial mask by
averaging each pixel over its $7\times7$ adjacent neighbours. We compound the
angular selection function as the set of all pixels with a value that exceeds a
given threshold. Finally, we construct a SDSS-DR7 mock catalogue from the
simulation box, by means of radial and angular selection functions. In
addition, a mock group sample is produced from the mock catalogue, by applying
the same percolation algorithm used for real groups (see section
\ref{sec:obser}). In our mock catalogue, the shapes of groups will be
calculated in the same way as in the observational data, i.e. using only the
projected positions of group members on the sky.

\subsection{Three-dimensional Results: anisotropic two point
            cross-correlation function}
\label{sec:3dcorr}

\begin{figure} 
\epsfxsize=0.5\textwidth 
\centerline{\epsffile{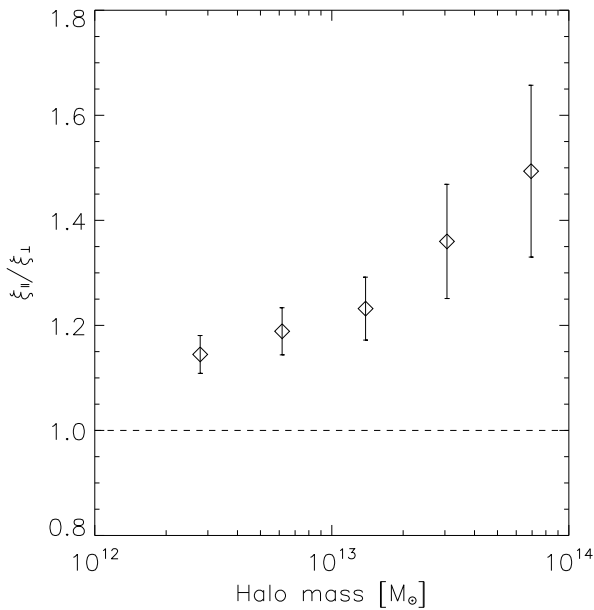}}
   \caption
   {
   Global ratios between three dimensional correlation functions along the
   directions parallel  and perpendicular to the central halo major axis
   ($\xi_{\parallel a}/\xi_{\bot a}$).  This ratio is computed by counting
   pairs with separations on the $2$-halo scale regime. The horizontal dashed
   line corresponds to a no anisotropy behaviour.
   }
\label{fig:binzo3D}
\end{figure}

We use the spatial halo-particle cross-correlation function $\xi(r)$ to study
the relation between the anisotropy presented by the distribution of matter and
the orientation of halo shapes.  This cross-correlation function measures the
probability excess to find a dark matter particle (tracer) in the volume
element $dV$ at a distance $r$ from a given halo (centre) \citep{Peebles}.  A
standard method to measure $\xi(r)$ in a simulation box consists in counting
halo-particle pairs in a given distance bin, and normalising this quantity by
the expected number of pairs for a homogeneous distribution. \citet{PaperL}
perform these counts restricting the pairs to fixed ranges in the angle between
their relative position vector and a given direction. In our case, since we
estimate the halo-particle cross correlation function, we restrict
halo-particle pair counts to different ranges of the angle subtended by its
relative position vector and the central halo shape axes (i.e. $\hat{a}$,
$\hat{b}$ or $\hat{c}$).  We will refer to the classical cross-correlation
function, computed using all pairs regardless of their angle as the
``isotropic'' cross-correlation function ($\xi_{\rm iso}$). We define the
``parallel'' cross-correlation functions as those computed using all pairs
subtending an angle with respect to a given halo shape axis smaller than a
given threshold $\theta_1=60^o$ (we justify this value later in this
section). These functions will be denoted $\xi_{\parallel a}$, $\xi_{\parallel
b}$ and $\xi_{\parallel c}$, for the corresponding axis $\hat{a}$, $\hat{b}$
and $\hat{c}$. 

We will also measure the dependence of the amplitude and shape of the parallel
cross-correlation functions on halo mass ($\rm{M_{halo}}$). For this purpose,
the halos in the simulation are split into five samples with different mass
ranges.  Figure \ref{fig:3D} shows in dotted lines the isotropic halo-particle
cross-correlation function $\xi_{\rm iso}$ for low mass systems
($\rm{M_{halo}}\leq 2.5\times10^{13} M_{\odot} h^{-1}$, left panel) and high
mass systems ($\rm{M_{halo}}\geq 1.3\times10^{14} M_{\odot} h^{-1}$, right
panel).  The short dashed, long dashed and solid lines correspond to the
cross-correlation functions estimated using particles in the direction parallel
to the minor, intermediate, and major axes of the halo shape, respectively, as
described above. The bottom panels show the ratio between the directional
cross-correlation functions and the isotropic function, with the same line
styles as in the upper panels. Shaded areas correspond to Jacknife error bands
(1--$\sigma$ uncertainties), estimated using the jacknife technique with
a total of $100$ subsamples for both, the numerical simulation and the SDSS
data. Our jacknife errors estimations are stable between $20$ to $150$
subsamples, without error underestimations.  We will apply this error estimate
throughout this paper.

The halo-particle cross-correlation functions in Figure \ref{fig:3D} show the
characteristic $1-$ and $2-$halo terms reported in several previous
measurements of both, observational \citep{Zehavi2004,2005Yang} and numerical
simulations \citep[see for instance][]{HayashiWhite2008}. The dashed
vertical line represents the transition scale between $1-$ and $2-$halo terms
(where the correlation function has a local minimum in the first derivative). 

In Figure \ref{fig:3D} it can be seen that the parallel cross-correlation
functions along the minor and intermediate halo shape axes ($\xi_{\parallel c}$
and $\xi_{\parallel b}$), show a lower amplitude than the isotropic classical
cross-correlation function ($\xi_{\rm iso}$), which in turn is smaller
than the cross-correlation function measured along the halo major axis
direction ($\xi_{\parallel a}$).  This measurement can be interpreted as an
excess of surrounding large scale structure along the direction of halo major
axis. The decrement of correlation power shown by $\xi_{\parallel c}$ and
$\xi_{\parallel b}$, could signal a decrement of the particle density in the
surrounding matter distribution in the plane perpendicular to the central halo
major axis. The drop in the correlation amplitude with respect to the isotropic
case observed for the $\xi_{\parallel b}$ function is smaller than that
observed for the $\xi_{\parallel c}$. This results reflects the triaxial nature
of the structure surrounding a typical halo. In spite of this quantitative
difference both, $\xi_{\parallel b}$ and $\xi_{\parallel c}$ exhibit a
decrement rather than an increment, that could be understood as a weak
preference for a triaxial over a prolate configuration for the surrounding
structure. For instance, if $\xi_{\parallel b}$ would showed an increment with
respect to the isotropic function in the same way as $\xi_{\parallel a}$ does,
we would have interpreted instead an oblate tendency for the surrounding
structure. It is also important to remark that  $\xi_{\parallel b}$ becomes
consistent with the isotropic function $\xi_{\rm iso}$ at large enough scales.

Given the prolate tendency shown by the cross-correlation function analysis, as
well as the similar amplitudes of $\xi_{\parallel b}$ and $\xi_{\rm iso}$, from
this point on, we will only study the anisotropy of the matter distribution
around dark matter halos using directions defined by the halo shape major axis.
In addition to this, we compute the ``perpendicular'' cross-correlation
function by adding pairs lying an angle $\theta_2$ or lower from the plane
perpendicular to the halo major axis direction. This function is denoted by
$\xi_{\bot a}$.  We choose the threshold angles, for the parallel and
perpendicular cases mentioned above ($\theta_1$ and $\theta_2$, respectively),
so that the volumes in each case are the same.  This can be achieved by
setting, $\sin(\theta_2) = 1 - \cos( \theta_1 ) =\chi$, and choosing a value
for the threshold parameter $\chi$. When different values of $\chi <0.5$
are chosen, the angles $\theta_1$ and $\theta_2$ tend to acquire more acute
values, while maintaining equal volumes.  When adopting different values of $\chi$
we found that the ratios between parallel and perpendicular correlations,
$\xi_{\parallel d} / \xi_{\bot d}$ (where $d$ represents either axis, $a$, $b$
or $c$) remain roughly constant, and the significance increases for larger
volumes, up to $\chi=0.5$.  When the volumes for the parallel and perpendicular
correlation are not the same, $\xi_{\bot d}$ and $\xi_{\parallel d}$ show
different degrees of departure from the isotropic correlation; in particular, the
correlation corresponding to the largest volume tends to be more similar to the
isotropic case. Consequently, we have selected $\chi=0.5$, which implies angles
$\theta_1 = 60^o$ and $\theta_2 = 30^o$. This choice ensures that every
halo-particle pair contributes to one of the two volumes.

The results for the perpendicular cross-correlation function are shown as
dot-dashed lines in Figure \ref{fig:3D}. As can be seen, its amplitude is quite
similar to the parallel cross-correlation function along the minor
axis. In fact, over most of the $2$-halo regime these two correlation
functions are indistinguishable according to the jacknife error bands.
Additionally, there is also an important difference between the parallel and
perpendicular cross-correlation functions (with respect to the major axis
of the halo shape, $\hat a$); their relative amplitudes show highly
statistically significant differences.  The transition between the $1-$ and
$2-$halo terms can be clearly seen in the cross-correlation function
quotients.  Given the halo shape definition (Equation \ref{eq:shape}) and the
aspherical geometry of halos \citep{PaperShapes}, it is expected that on
scales within the $1-$halo term, the parallel and perpendicular
cross-correlation functions will present a significant difference, since they
will simply mirror the intrinsic halo shapes.  A more relevant result is that
coming from the comparison between the cross-correlation functions over
the $2-$halo term regime, which shows an approximately constant behaviour and
only loses statistical significance at very large separations. Figure
\ref{fig:3D} also shows that the anisotropy of the cross-correlation
function for the high mass sample (right panel) is higher than for the low mass
sample. We have analysed five samples with halo masses ranging
from $10^{12}M_{\odot} h^{-1}$ to $2\times10^{14}M_{\odot} h^{-1}$,
confirming that the 2-halo term alignment signal increases with the
halo mass.

As is described above, the anisotropy signal behaviour is approximately constant
over the $2-$halo clustering regime. Therefore, it is possible to characterise
a median alignment signal over this range by counting all pairs with
separations in the $2-$halo regime, in order to estimate a global quotient
between correlation functions along the directions parallel and perpendicular
to the halo centre major axis. These results are shown on Figure
\ref{fig:binzo3D} for five samples constructed using different halo centre mass
intervals. The median mass for each sample is shown on the abscissas, whereas
the ordinate shows the global ratio $\xi_{\parallel a}/\xi_{\bot a}$.  Error
bars are estimated using the jacknife resampling technique. As can be seen in
the figure, there is an increment of the degree of anisotropy as the mass
increases.  This is expected in the hierarchical halo formation scenario since
more massive halos have more recent formation epochs and therefore they are
more likely to be aligned with the present large scale matter distribution.

\subsection{Projected anisotropic cross-correlation function and
            resolution effects}
\label{sec:projcorr}

The analysis of observational data from redshift surveys is mainly
limited by three important effects.  (i) The measurement of galaxy group
shapes needs to be done in projection on the plane of the sky due to the
presence of galaxy peculiar velocities which affect distance estimations based
on redshift measurements. In order to asses the reliability of the estimates of
projected group shape orientations, it is necessary to reproduce this effect in
the simulation, and to study whether the alignment signals can still be
detected in this case.  The group shape tensor will therefore be calculated
using only two directions along the plane of the sky. Due to the flux-limited
nature of galaxy catalogues, (ii) only a few group members are available to
estimate the shape tensor, and (iii) the number of tracers is diluted as the
distance increases (both effects become stronger at higher redshifts). The
decrement of galaxy group members (ii) tends to produce deviations of the
estimated group shape orientation with respect to the real direction. This is a
shot noise type effect which also biases the measured shapes towards oblate
spheroids \citep{PaperShapes}. The dilution of the number density of galaxies
with redshift can be taken into account with a proper modelling of the selection
function of the catalogue via, for instance, constructing mock catalogues.  We must
estimate the influence of all these effects before proceeding further with the
analysis of the real data.

We start by projecting the simulated volume onto the Cartesian plane $x-y$ in
the numerical simulation.  This plane resembles the ``plane of the sky'' of the
observational data.  The third Cartesian coordinate $z$ plus the $z$-component
of the peculiar velocity, is used to estimate a radial velocity along the
``line of sight''.  This last quantity mimics the redshift measured in
observations.  The effects of low number statistics on the measured group
shapes can be taken into account by selecting at random a small number of
particle members in order to compute the projected shapes. Even though galaxies
in groups do not necessarily follow the dark matter halo density profile,
\citet{PaperShapes} have shown that this procedure reproduces the biases
present in the observational estimates of group shapes due to projection and
shot noise effects, that is, the shapes of simulated dark matter halos measured
this way exhibit indistinguishable differences with shapes obtained from galaxy
groups in both, real and mock catalogues. 

We now perform the equivalent of the analysis presented in section
\ref{sec:3dcorr}, using  anisotropic projected two-point cross-correlation
functions.  In an analogous manner to the method implemented on the
three-dimensional case, we separate halo-particle pairs according to the angle
subtended with respect to the apparent halo major axis.  Pairs with angles
greater than and less than a threshold angle of $\theta_p = 45^o$, define the
parallel ($\omega_{\parallel}$) and perpendicular ($\omega_{\bot}$) projected
cross-correlation functions respectively. Notice that this choice of
$\theta_p$ is analogous to that of $\theta_1$ and $\theta_2$ for the three
dimensional case, since this separates the plane of the sky in two equal areas.
The quotients $\omega_{\parallel}/\omega_{\bot}$, for smaller fractions of the
area of the sky are roughly constant with a decreasing statistical significance
(due to a lower number of counts). This behaviour is analogous to the observed
in the three dimensional signal. As a means to avoid a large amount of
foreground and background structures we only consider neighbours in distance
slices of $50$ Mpc h$^{-1}$ around the halo centres.  Differences between the
correlation amplitudes of $\omega_{\parallel}$ and $\omega_{\bot}$ will then be
interpreted as statistical excesses of the surrounding large scale structure
along the direction of the central group projected major axis. This will
represent a lower limit of the real anisotropy around the group major axis
given the averaging of the orientation of the 3D shape major axis with respect
to the line of sight. 

\begin{figure}
\epsfxsize=0.5\textwidth 
\centerline{\epsffile{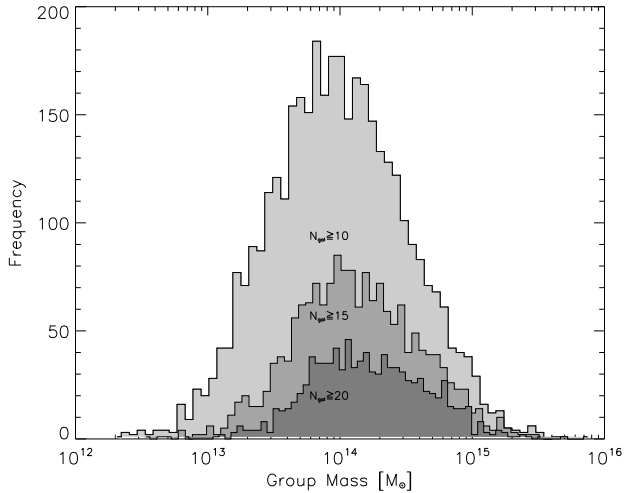}}
\caption
     {
     Distribution of group masses for different samples selected according to
     minimum thresholds on number of galaxy members. The lighter shaded
     histogram shows the results for the main galaxy group sample.
     Distributions of galaxy group subsamples with more than $15$ and $20$
     galaxy members are shown with darker shaded areas as indicated in the
     figure key.
     }
     \label{fig:mass}
\end{figure}

With the purpose of studying the dependence of the anisotropy with the mass, we
will measure the cross-correlation functions selecting halo centres separated
into three mass ranges corresponding to the terciles of the Sloan group mass
distribution (see section \ref{sec:obser} for the description of the group
sample).  This choice allows a closer comparison with the observational data,
and ensures the same number of galaxy groups in each mass range in the
observational analysis (see section \ref{sec:obsercorr}).  On the other hand,
it is well known that the estimate of a halo shape depends on the number of
members used to compute it \citep{PaperShapes}. Therefore, it is expected that
this will affect the alignment signal obtained from the ratios between
anisotropic and isotropic cross-correlation functions. In order to analyse the
extent of this effect we compute the halo shapes using only $10$, $15$ and $20$
random particles from each halo. Furthermore, to better mimic observations we
impose that the distributions of halo mass resemble the real galaxy group
distributions with $10$, $15$ and $20$ members, respectively (see Figure
\ref{fig:mass}). 

\begin{figure}
\epsfxsize=0.454\textwidth 
\centerline{\epsffile{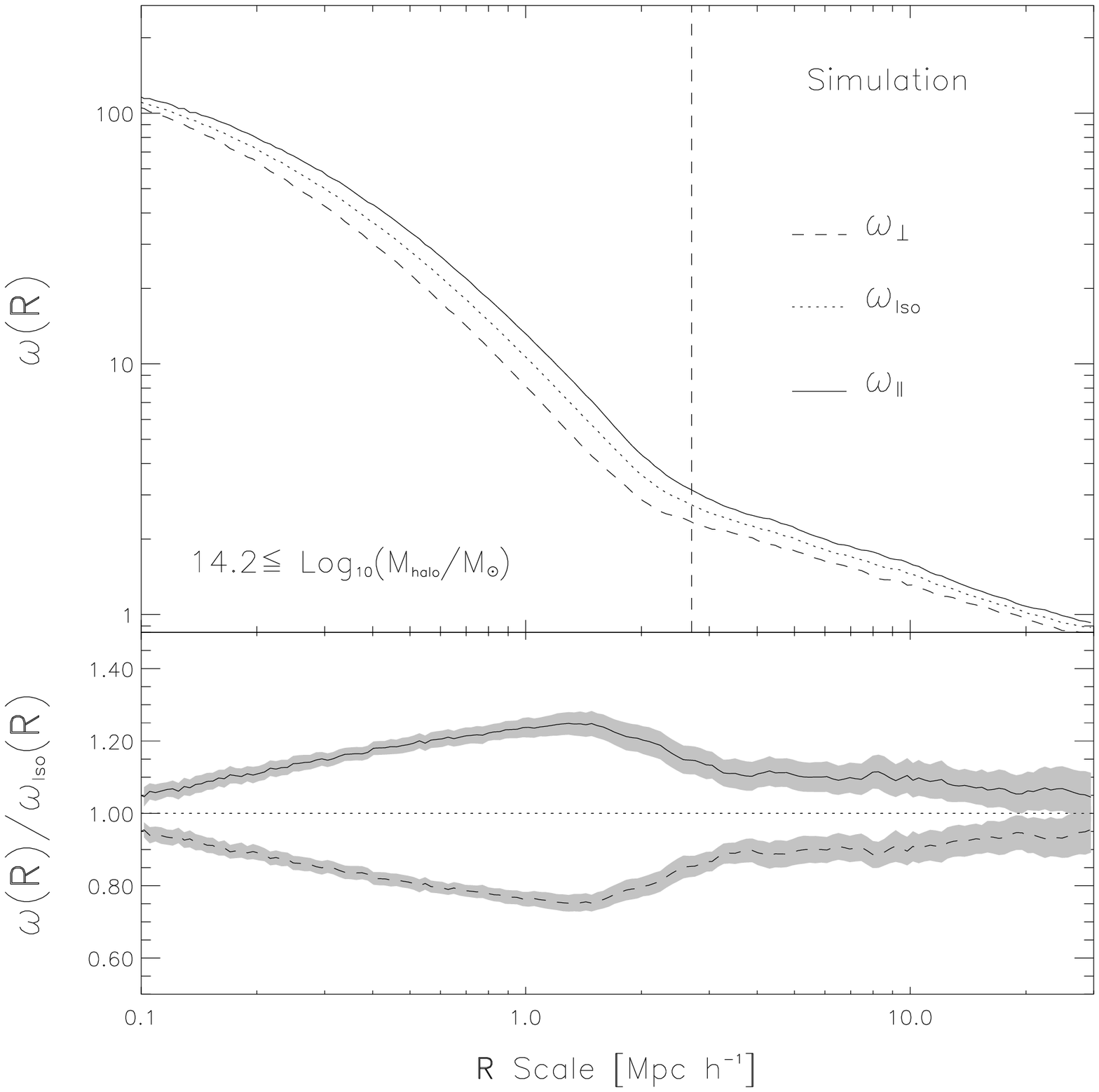}}
\caption
     {
     Anisotropic projected correlation functions for massive halos (third
     tercile) with a major axis direction determined using only $10$ randomly
     selected particles. The mass distribution was forced to resemble the
     corresponding real galaxy group mass distributions with $10$ members.
     }
\label{fig:baqueteado}
\end{figure}

Figure \ref{fig:baqueteado} shows the anisotropic projected halo-particle
cross-correlation functions for the third tercile on the mass of groups with at
least $10$ members ($\rm{M_{halo}}\geq 1.7\times10^{14}M_{\odot} h^{-1}$). The
halo major axis direction was determined using $10$ particles selected at
random. Solid, dotted and dashed lines correspond to parallel, isotropic and
perpendicular cross-correlation functions to major axis, respectively. The
lower panels show the quotients between anisotropic and isotropic
cross-correlation functions.  By comparing this figure with right panel of
Figure \ref{fig:3D} it can be seen that the alignment signal decreases by
approximately a $75\%$ due to shot noise and projection effects. It can also be
seen that the quality of the signal decreases, lowering the statistical
significance by some degree.  However, the detection of the alignments remains
strong.  The error bars decrease significantly when $20$ particle members are
used in the estimate of the major axis direction.

So far, we have studied the combined effect of projection and shot noise; a
third effect mentioned at the beginning of this section, the flux-limited
nature of observational catalogues, needs to be analysed.  To this end, we use
artificial data sets that resembles the real SDSS-DR7 catalogue (see Section
2.1) and the real galaxy group catalogue.

\begin{figure*} 
\centering
\begin{picture}(430,465)
\put(-20,240){\psfig{file=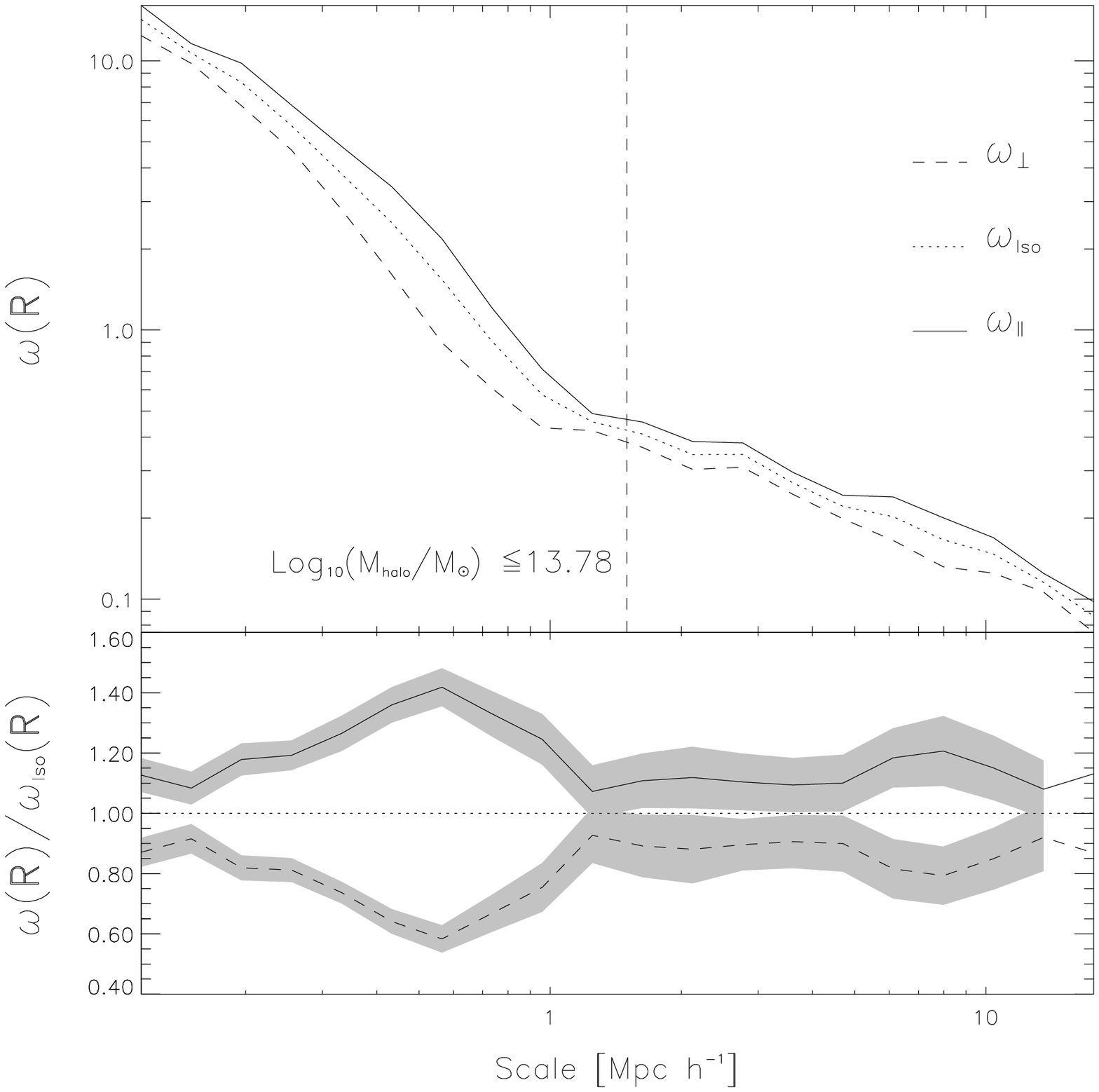,width=8.5cm}}
\put(215,240){\psfig{file=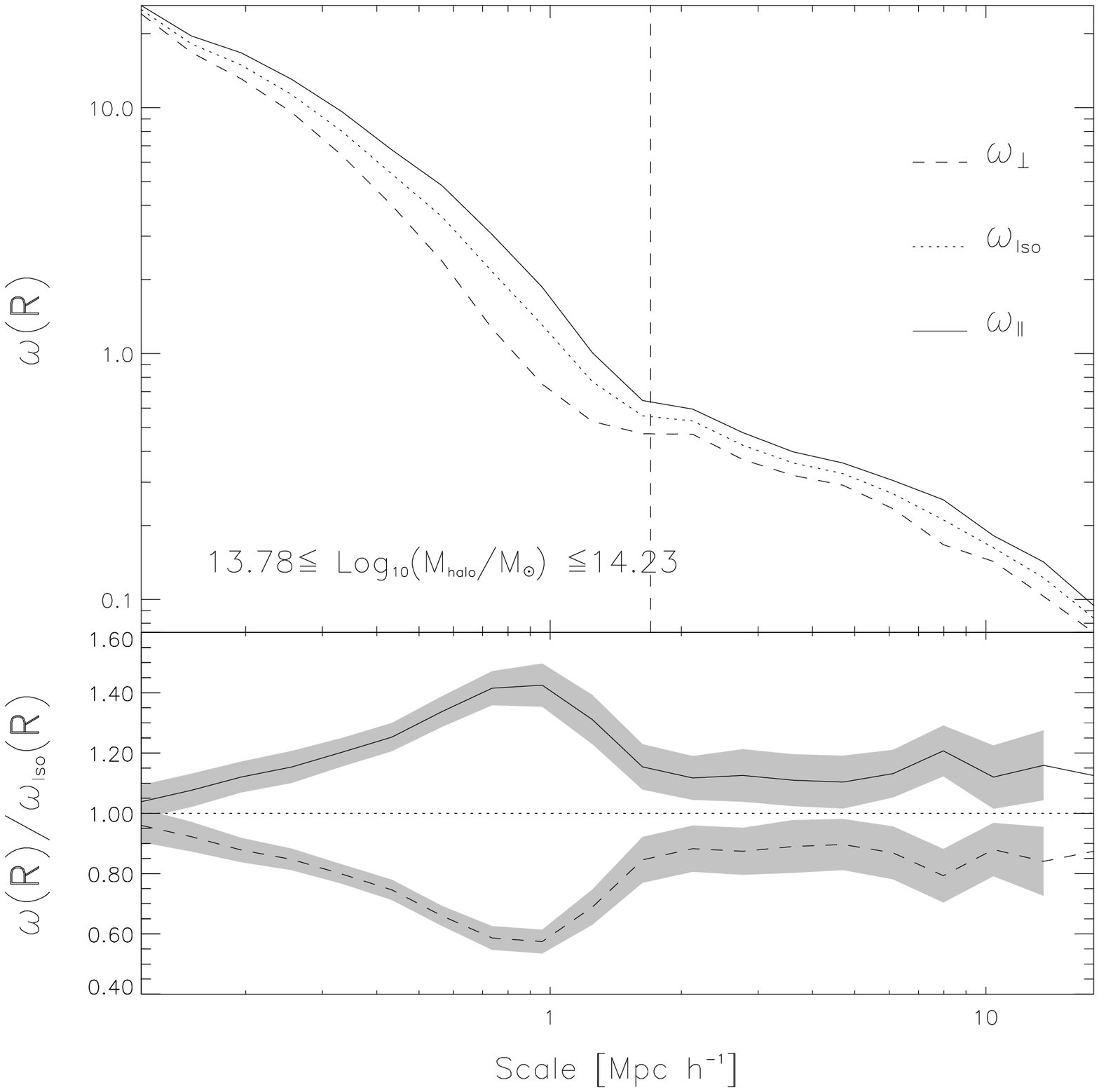,width=8.5cm}}
\put(-11,10) {\psfig{file=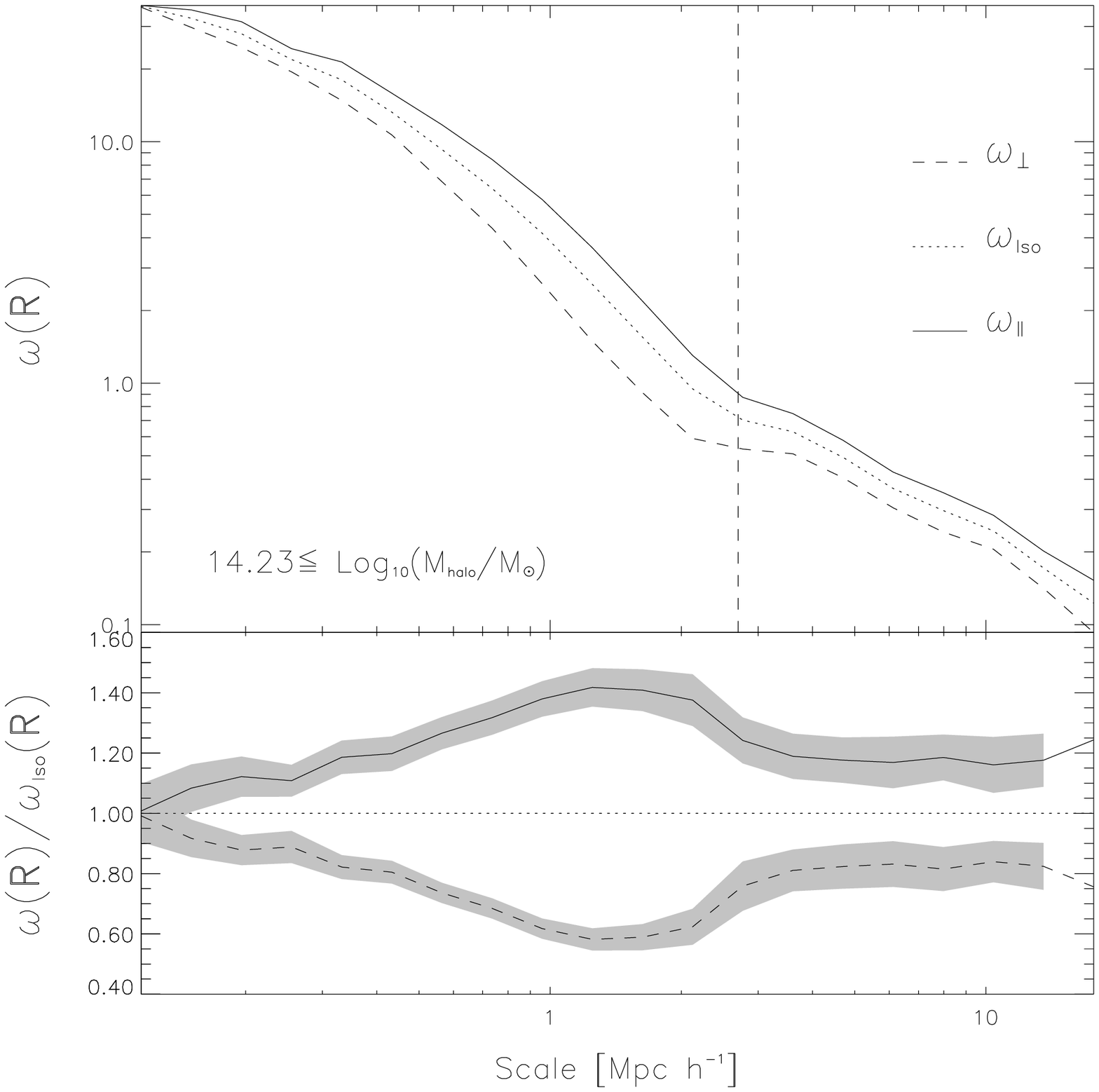,width=7.69cm}}
\put(224,10) {\psfig{file=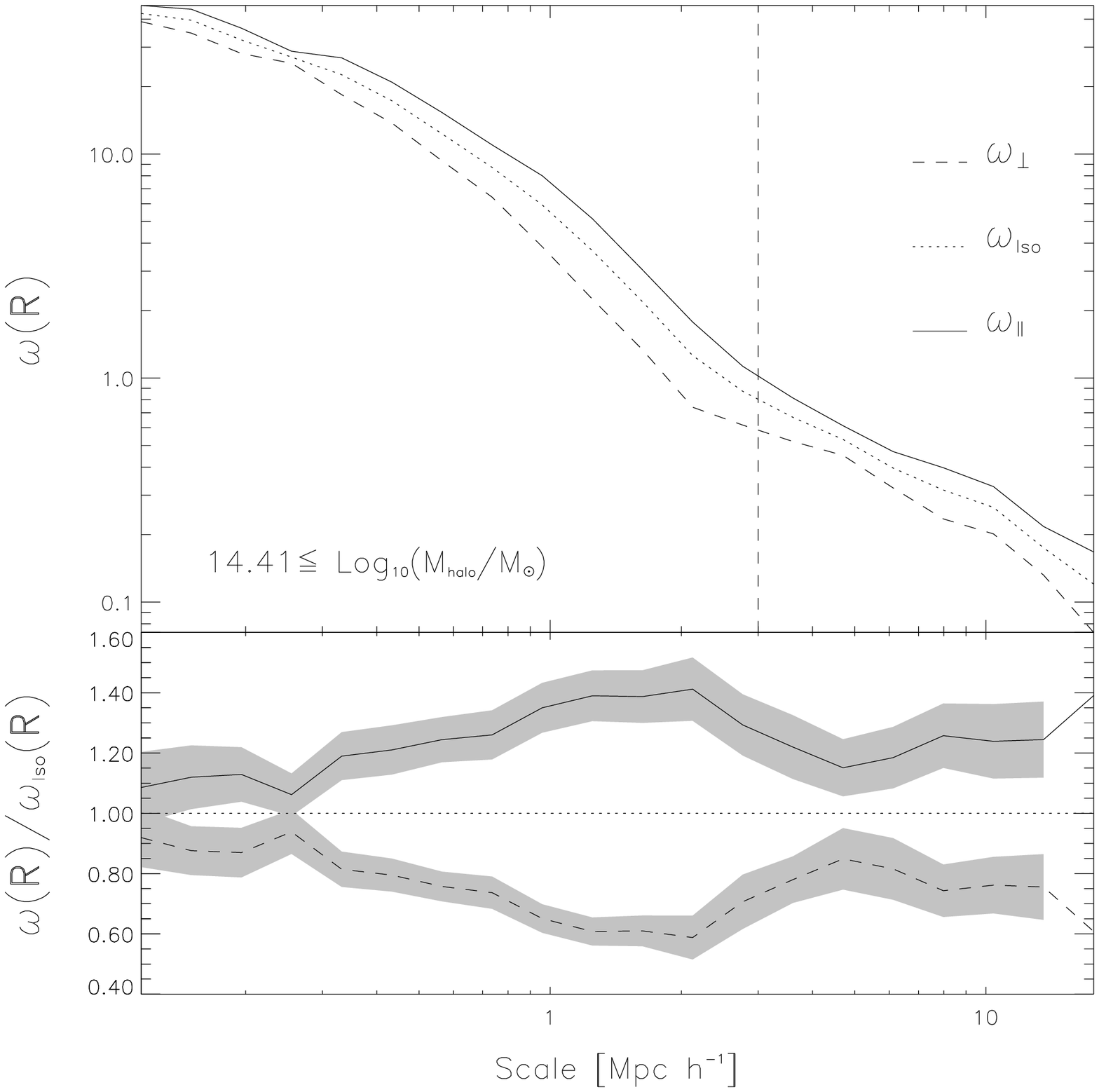,width=7.69cm}}
\put(25,445){\Large{Mock data}}
\put(180,445){\Large{(A)}}
\put(415,445){\Large{(B)}}
\put(180,210){\Large{(C)}}
\put(415,210){\Large{(D)}}
\end{picture}
       \caption
       {
       Projected group$-$galaxy correlation functions for three group mass
       ranges extracted from a mock SDSS catalogue. The dashed lines correspond
       to the correlation function between mock groups and mock neighbour
       galaxies in the direction perpendicular to the group shape major axis,
       as seen projected onto the sky.  The solid lines show the results when
       using tracers along the direction of the projected major axis, and the
       dotted lines show the corresponding results when all neighbours are taken
       into account.  The first three panels, A, B and C, show the results for
       the lowest, intermediate and highest mass terciles of the mock galaxy
       group sample selected with at least $10$ mock galaxy members per group.
       The lower right panel, labelled D, shows the results for the high mass
       tercile with at least $20$ members ($M\ge 2.6\times 10^{14}\,$h$^{-1}
       M_\odot$).  The ratios between the correlation functions along different
       directions and the correlation function obtained using all the neighbours
       are shown in the lower subpanels.
       }
\label{fig:mock}
\end{figure*}

Based on this mock data we compute again the parallel ($\omega_{\parallel}$)
and perpendicular ($\omega_{\bot}$) projected group-galaxy cross-correlation
functions, by applying the same procedure used on the observations, as will be
described in the following sections (see Section \ref{sec:obsercorr}). The
results are shown in Figure \ref{fig:mock}.  Line types are as in Figure
\ref{fig:baqueteado}. The first three panels (A, B and C) show the projected
anisotropic cross-correlation functions for different mass terciles, for mock
groups with at least $10$ galaxy members. In the lower right panel of Figure
\ref{fig:mock} (panel D) we show the results for mock groups in the highest
mass tercile with at least $20$ group members.

We now compare the results from projected and three dimensional
correlation functions with the aim to assess the degree of loss of the
alignment signal of section \ref{sec:3dcorr}. The analysis of the expected
biases allows us to conclude that the alignment signal obtained in section
\ref{sec:3dcorr} is still detectable on projected measurements of the two point
cross-correlation function.  When only considering the effect of projection we
find, for scales around $10$ Mpc h$^{-1}$, differences between correlation
quotients in the directions parallel and perpendicular to the halo orientation
decrease by a $50$\%, with respect to the corresponding three-dimensional ratio
(Figure \ref{fig:3D} right panel).  When the mass distribution of observational
galaxy groups is reproduced (Figure \ref{fig:mass}), and also the discreteness
effect on the determination of group shapes is taken into account (Figure
\ref{fig:baqueteado}), the anisotropy signal diminishes even more, another
$50$\% at scales around $10$ Mpc h$^{-1}$, that is a $75$\% with respect to the
quotients shown in Figure \ref{fig:3D}.  Finally, when the depth and geometry
of the observational survey are taken into account, as well as the galaxy group
identification is mimic, the error bars increase significantly.  However, the
anisotropy signal remains detectable with a high statistical significance over
a wide range of distances.

\section[]{Observed correlation alignments on SDSS real data}

In this section we analyse the presence of galaxy group shape alignments with
the large scale structure as traced by real galaxies. To this end, measurements
of anisotropic projected cross-correlation functions were performed searching
for an alignment signal in the $1-$halo and $2-$halo clustering regimes.  At
the end of this Section we provide a comparison between observational and
numerical simulation results.

\subsection{The galaxy group sample} \label{sec:obser}

We use the main galaxy sample from the 7th Data Release of the Sloan Digital
Sky Survey \citep[SDSS-DR7,][]{Abazajian_DR7} which includes roughly
$\approx900,000$ galaxies with redshift measurements to $z\approx0.3$ and an
upper apparent magnitude limit of $17.77$ in the r band. The galaxy groups used
in this work have been identified in this sample using the same procedure as in
\citet{MerchanGRP}. The method consists in using a friend-of-friend algorithm
similar to the one developed by \citet{1982Huchra}.  The algorithm links pairs
of galaxies $(i, j)$ which satisfy $D_{ij} \leq D_0 R(z)$ and $V_{ij} \leq V_0
R(z)$ where $D_{ij}$ is the projected distance and $V_{ij}$, the line-of-sight
velocity difference. The scaling factor $R(z)\propto {\rm n}_{\rm g}^{-3}$
is introduced in order to take into account the decrement of the observed
galaxy number density, ${\rm n}_{\rm g}$, with redshift due to the apparent
magnitude limit cutoff \citep[see eq. $5$ in][]{1982Huchra}. We have adopted a
transverse linking length $D_0$ corresponding to an overdensity of
$\delta\rho/\rho = 80$, a line-of-sight linking length of $V_0 = 200$ km
s$^{-1}$ and a fiducial distance of $10$ Mpc h$^{-1}$.  Our main group sample
comprises all galaxy groups with at least $10$ galaxy members. Group virial
masses are estimated following \citet{MerchanGRP,1960LM},
$$ {\rm M}=\frac{\sigma^2 R_v}{G}, $$
where $\sigma$ is the three dimensional velocity dispersion, $R_v$ is the
virial radius \citep[see eq. $6$ in][]{MerchanGRP} and $G$ is the gravitational
constant. The velocity dispersion $\sigma$ is estimated using its
observational counterpart, the line-of-sight velocity dispersion $\sigma_v$,
$\sigma = \sqrt{3}\sigma_v$.

Figure \ref{fig:mass} shows the mass distribution for the main group sample and
for two additional subsamples with at least $15$ and $20$ galaxy members. In
order to analyse the dependence of any alignments with mass, we divide the
ordered mass distribution in three parts, each containing a third of the sample
population. The halo mass distribution shifts and stretches to higher masses
when a higher minimum number of galaxy members is imposed. 

\subsection{Observed alignments on the projected cross-correlation function}
\label{sec:obsercorr}

\begin{figure*} 
\centering
\begin{picture}(430,465)
\put(-20,240){\psfig{file=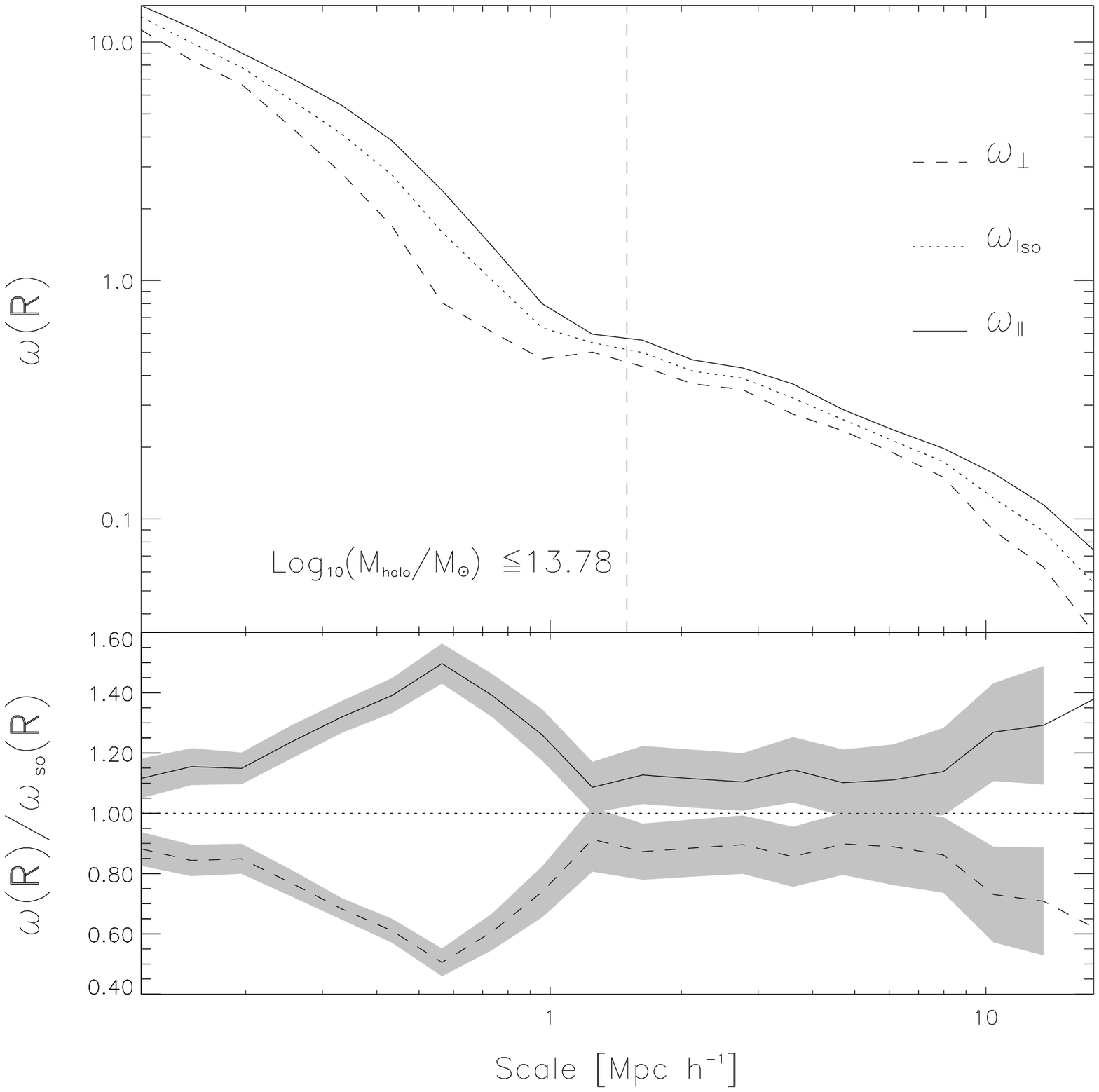,width=8.5cm}}
\put(215,240){\psfig{file=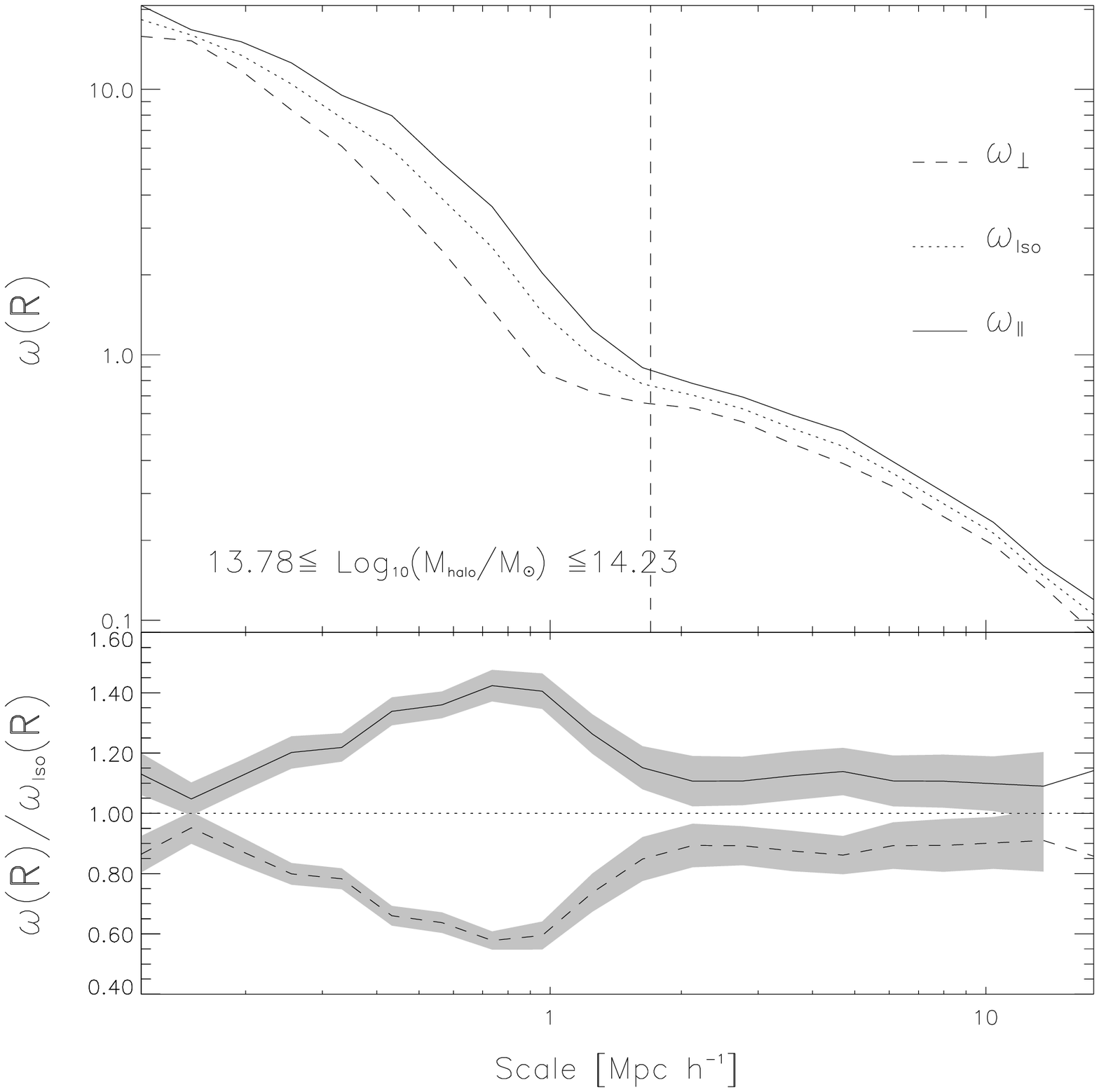,width=8.5cm}}
\put(-11,10) {\psfig{file=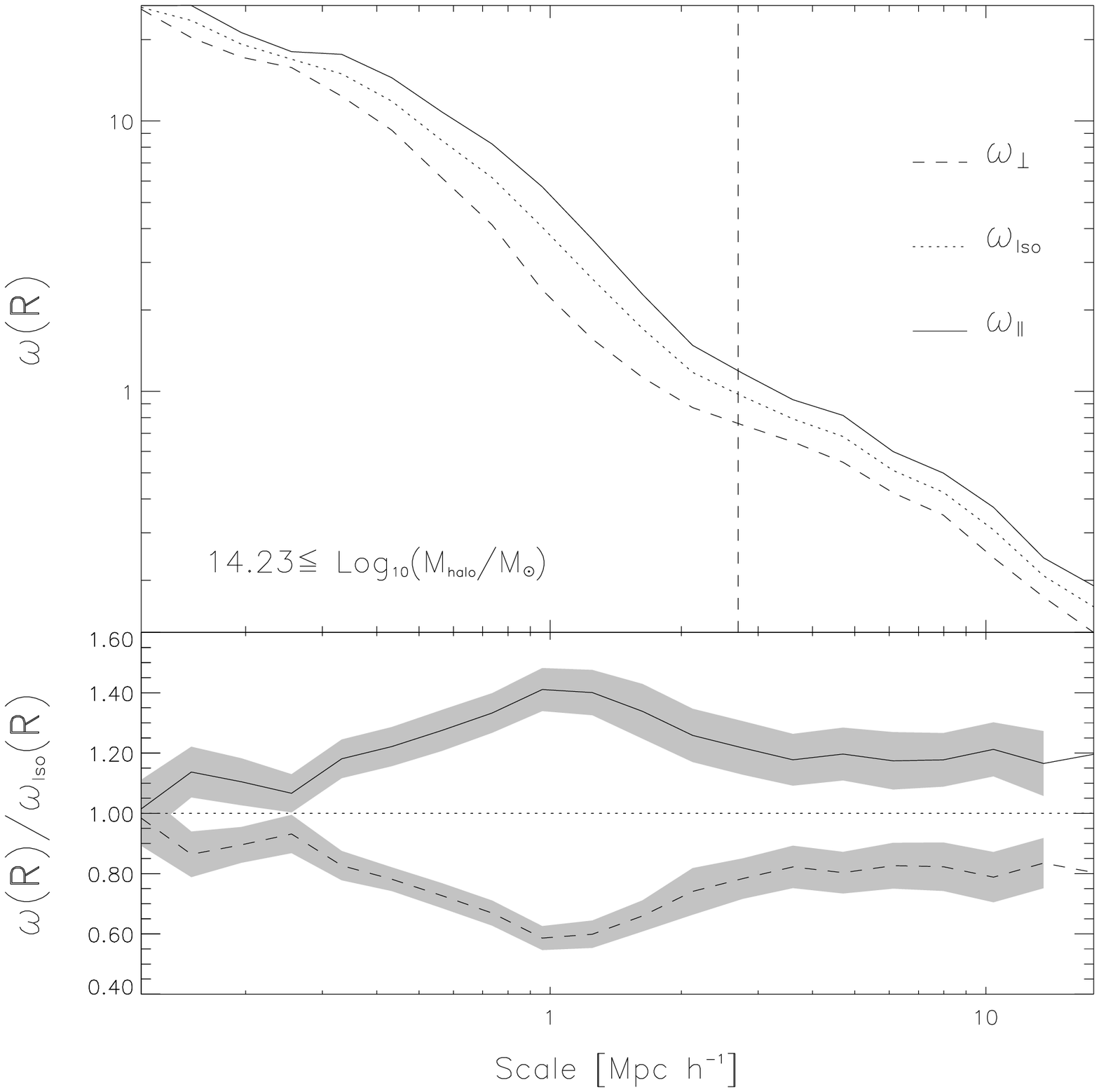,width=7.69cm}}
\put(224,10) {\psfig{file=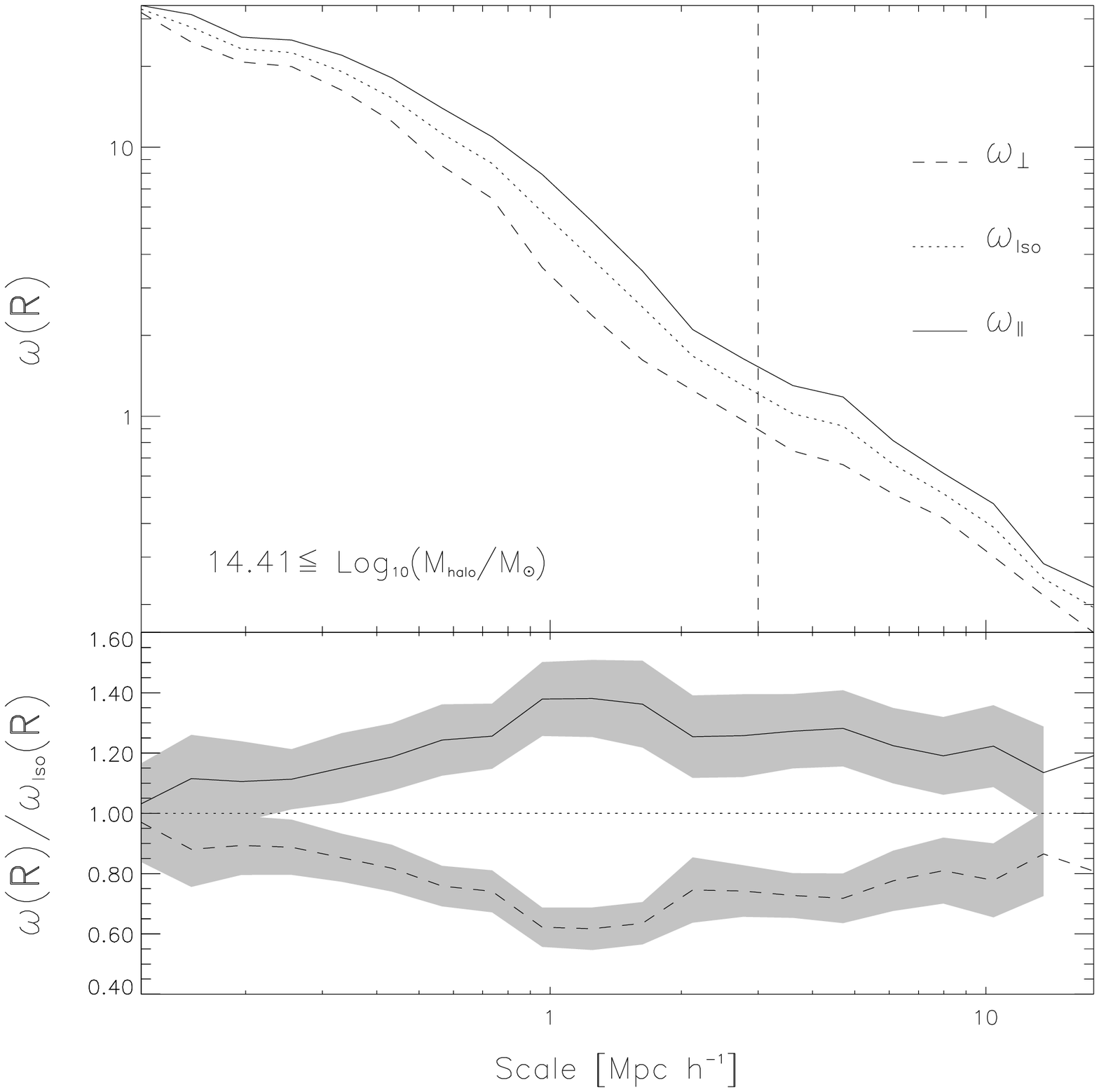,width=7.69cm}}
\put(25,445){\Large{Real data}}
\put(180,445){\Large{(A)}}
\put(415,445){\Large{(B)}}
\put(180,210){\Large{(C)}}
\put(415,210){\Large{(D)}}
\end{picture}
       \caption
       {
       SDSS projected group$-$galaxy correlation functions for three group mass
       ranges.  The dashed lines correspond to the correlation function between
       groups and neighbour galaxies in the direction perpendicular to the group
       shape major axis, as seen projected onto the sky.  The solid lines show
       the results when using tracers along the direction of the projected
       major axis, and the dotted lines show the corresponding results when all
       neighbours are taken into account.  The first three panels, A, B and C,
       show the results for the lowest, intermediate and highest mass terciles
       of the galaxy group sample selected with at least $10$ galaxy members
       per group.  The lower right panel, labelled D, shows the results for the
       high mass tercile with at least $20$ galaxy members ($M\ge 2.6\times
       10^{14}\,$h$^{-1} M_\odot$).  The ratios between the correlation
       functions along different directions and the correlation function
       obtained using all the neighbours are shown in the lower subpanels.
       }
\label{fig:obseres}
\end{figure*}

In section \ref{sec:3dcorr} the correlation function measured in the simulation
was estimated by counting halo-particle pairs and normalising by the expected
number of pairs for a homogeneous distribution. Due to the complex selection
function of the SDSS survey, it is necessary to produce a random distribution
of points with the same selection function in order to normalise the
group-galaxy pair counts. This normalisation can be done in several ways; in
this work we use two estimators, the classic estimator \citep{Davistem}
$\omega=DD/DR-1$, where $DD$ and $DR$ are the numbers of {\it group-galaxy} and
{\it group-random } pair counts respectively, and a symmetric version of the
\cite{LSestim} estimator; in all cases we find negligible differences between
these two estimators.  We perform all the analysis using the
\citeauthor{LSestim} estimator.  This computation is performed as
$\omega=(DD-DR-RD+RR)/RR$, where in addition to $DD$ and $DR$, we need to
calculate $RR$ and $RD$, the numbers of {\it random centre-random tracer} and
{\it random centre-galaxy} pair counts respectively. For each sample of groups
we compute a random centre catalogue that mimic its redshift distribution. The
number of random centres is about $100$ times the number of groups in each
sample. The homogeneous sample of tracers was computed from the expected
numerical density of galaxies at a given redshift, with a magnitude below the
limit of the survey, and a Schechter luminosity distribution with parameters
$\phi_*=0.0149 \;, M_*=-20.44\;, \alpha=-1.05$ \citep{BLum}. These random
catalogues contain $2\times10^7$ random points.  The angular positions are
restricted to the angular mask described in section \ref{sec:projcorr}.

In order to estimate the galaxy group anisotropic cross-correlation function we
follow the same definitions given in section \ref{sec:projcorr} (i.e.
$\omega_{\parallel}$, $\omega_{\bot}$, $\omega_{\rm iso}$). We therefore count
$DD$ and $DR$ pairs depending on their relative orientation with the major
projected axis of the group shape tensor. These counts are binned over a wide
range of scales covering from $0.1$ to $50$ Mpc h$^{-1}$. In analogy with the
estimate of the projected cross-correlation function in simulations (section
\ref{sec:projcorr}), we only consider galaxy neighbours with relative comoving
distance $<50$ Mpc h$^{-1}$ around the group centre.

In Figure \ref{fig:obseres} we show the resulting cross-correlation function
measurements.  Line types are as in Figure \ref{fig:baqueteado}.  The first
three panels (A, B and C) show the projected anisotropic cross-correlation
functions for different mass terciles, for groups with at least $10$ galaxy
members. In order to study the effect of low number statistics, we repeat the
same analysis for samples with more than $15$ and $20$ galaxy members.  For
simplicity only the results from the highest mass tercile are shown in panel D
of Figure \ref{fig:obseres}, for groups with at least $20$ galaxy members.

The first three panels in Figure \ref{fig:obseres}, show a similar behaviour to
that seen in the numerical simulation, where the anisotropy signal increases
with mass, particularly for the $2$-halo regime. In other words, significant
statistical differences are seen between the $\omega_{\parallel}$ and
$\omega_{\bot}$ projected cross-correlation functions, which become more
prominent as the mass of the centre group increases.  The number of galaxy
groups used to compute the results showed in panel D is approximately a third
of the total number of groups. Consequently, the correlation function errors
increase. Nevertheless, the signal is still significant.  A striking similarity
of the results on this figure with the corresponding mock results can be
observed by direct comparison with Figure \ref{fig:mock}.

The results shown in this section confirm the presence of alignments in the
large-scale structure as traced by galaxies and the direction of the major axis
of galaxy groups, in qualitative agreement with the expectations from
$\Lambda$CDM model. This alignment signal persist over the two halo regime, up
to scales of $20$ Mpc $h^{-1}$, pointing an alignment between the group inner
galaxy distribution and the surrounding large scale structure.

It should also be mentioned that the covariance matrix of the correlation
function has non-zero off-diagonal elements. In order to reduce the effects of
covariance between measurements at different scales, we estimated global ratios
between cross-correlation functions in the directions perpendicular and
parallel to the group major axes on two different wide scale ranges,
corresponding to the $1-$ and $2-$halo terms, characterised by projected
distances lower and greater than a transition scale. This procedure allows to
diminish the effect of error correlations on the anisotropy, and also allows to
study the alignments on these two different scale regimes. The scale threshold
depends on mass, varying from $0.5$ to $3$ Mpc h$^{-1}$ over the mass range
considered here.  

\begin{figure*} 
\begin{picture}(430,250)
\put(-40,0){\psfig{file=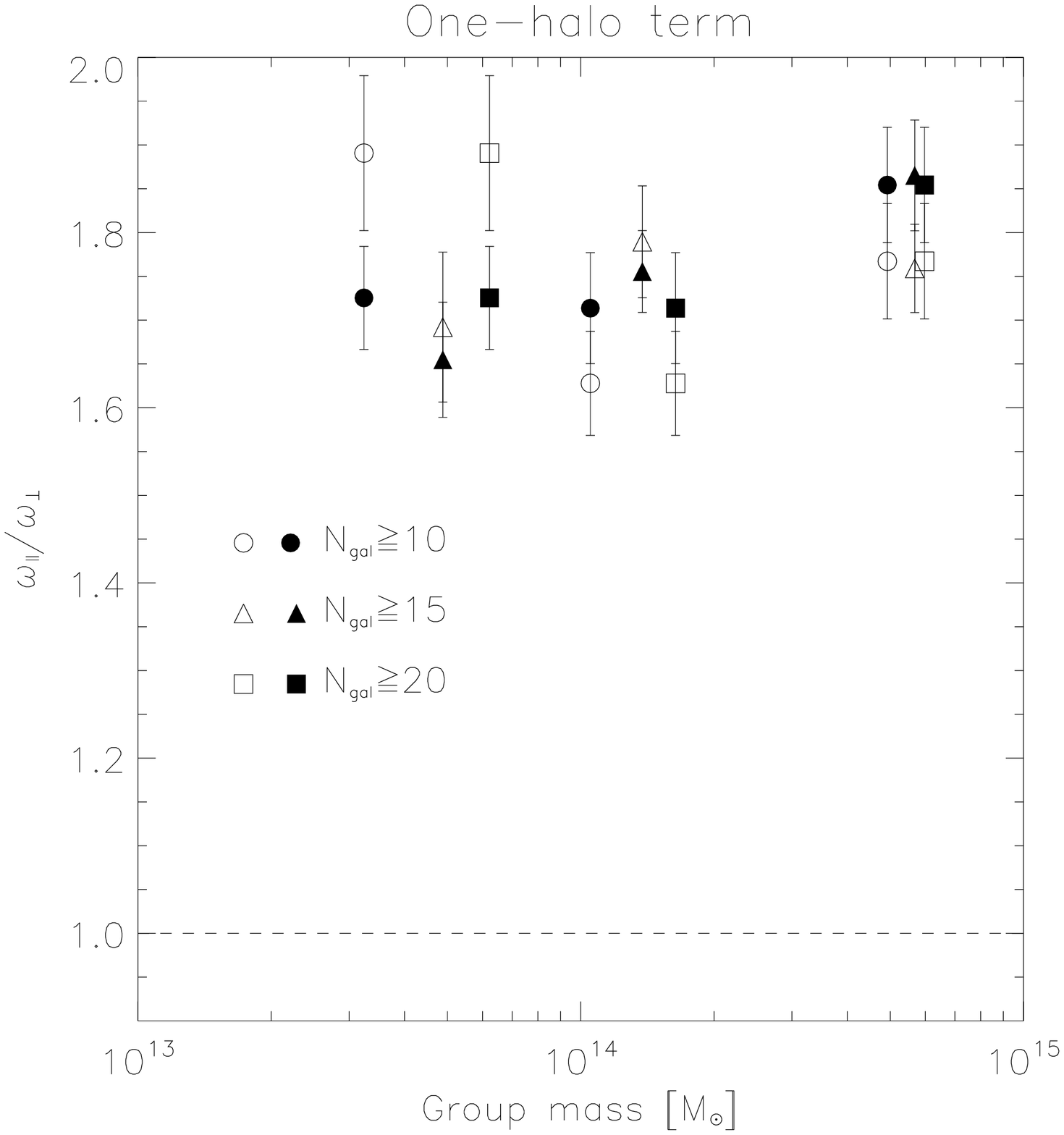,width=9.cm}}
\put(210,0){\psfig{file=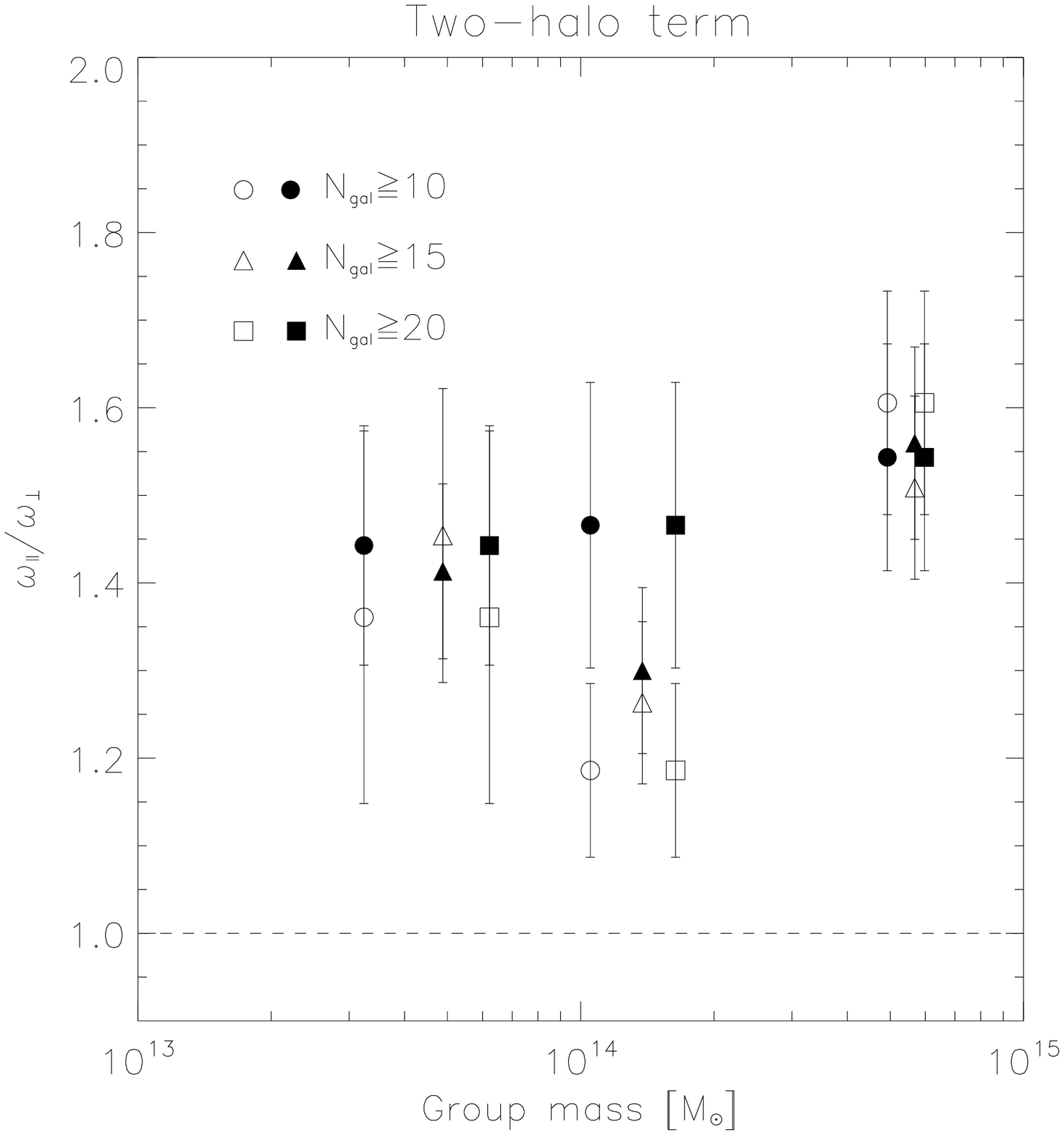,width=9.cm}}
\end{picture}
\caption
   {
   Global ratios between correlation functions along the directions parallel
   and perpendicular to the central group projected major axis
   ($\omega_\parallel/\omega_\perp$).  Two different ratios are computed by
   counting pairs split into the $1$- and $2$-halo clustering regimes (left and
   right panels respectively).  The open symbols represent the results obtained
   from real observational data.  Filled symbols show the corresponding results
   obtained from the SDSS mock catalogue.  The horizontal dashed lines
   indicate the unit ratio (no anisotropy).
   }
   \label{fig:compa2}
\end{figure*}

Figure \ref{fig:compa2} shows the $1-$ and $2-$halo term ratios as a function
of the average subsample mass (left and right panels, respectively).  In the
figure the open symbols  represent the results obtained from real observational
data, whereas filled symbols correspond to mock data described in section
\ref{sec:projcorr}. As shown in the figure key, the different open (filled)
symbols represent measurements obtained from real (mock) group samples with at
least $10$, $15$ and $20$ galaxy group members. This figure can be used to
analyse the dependence of the anisotropy on the group mass for the $1-$ and
$2-$halo regimes in real and simulated data.

The $1-$halo term ratios obtained from the numerical simulation reflect a
slight increase in the asphericity of halos as their mass increases, which is
qualitatively consistent with previous results on the projected shapes of
galaxy groups \citet{PaperShapes}. The observations do not show a clear
tendency, although error bars do not allow a clear distinction between mock and
real data, indicating that the use of the shape tensor is more adequate for the
purposes of studying the variation of the intrinsic shape of groups as a
function of mass. In most cases the observations are in agreement with the
numerical simulation results.

On the $2-$halo regime, the ratios shows a marginal increase with the group
mass, both in the numerical simulation and observational results, which do not
show a systematic dependence on the minimum number of group members used.
Within the uncertainties, the observational results are in agreement with mock
data. It can be seen for all mass samples, regardless of the number of group
members, a significant degree of anisotropy in the correlation ratios. For the
more massive sample this anisotropy is distinguishable from isotropy at a level
above $3$ standard errors.

\section{Conclusions}

We have used a technique developed by \citet{PaperL} to characterise the
alignment between the shape of gravitationally bound systems and the large
scale structure on a  $\Lambda$CDM numerical simulation and observed galaxy
groups taken from the Sloan Digital Sky Survey, Data Release 7.  We have
detected a statistically significant alignment of galaxy groups with the
surrounding large scale structure traced by SDSS spectroscopic galaxies. This
result, based on an up-to-date sample of spectroscopic galaxy groups, supports
the well known theoretical alignment of dark matter halo shapes with the
circumventing large scale structure.

The analysis on the $\Lambda$CDM numerical simulation shows that the direction
of maximum elongation of dark matter halos is aligned with the surrounding
matter distribution on scales of up to $30$ Mpc h$^{-1}$. The resulting
halo-particle cross-correlation function shows the characteristic transition
between the $1-$ and $2-$halo terms regimes which is also reflected in the
anisotropy signal. Given that the $1-$halo term corresponds to the inner matter
distribution of the halos, the alignment on this range of scales was expected.
More interesting to consider is the signal observed over the $2-$halo term
regime, which could be related with the halo formation history. It has been
established that halo formation occurs by anisotropic mergers with lower mass
halos. These accretion events occur along directions traced to some extent by
the large scale structure. Therefore, the existence of a remnant correlation
between halo shape orientations and the large scale matter distribution is
expected.  This constitutes a fossil record of the halo formation process.  Due
to the hierarchical nature of halo formation, an increasing anisotropy signal
as the halo mass increases is also expected, as more massive halos finished
assembling their mass more recently and therefore they are more likely to bear
marks of more recent important formation processes.  This mechanism also
explains the shape-mass relation observed in the SDSS galaxy group sample
\citet{PaperShapes}.

Our measurements of this alignment on the two different halo model regimes are
qualitatively consistent with previous results on dark matter halo alignments
in $\Lambda$CDM numerical studies \citep{2007Aragon_Spin, Hahn1,
2007Brunino_Shape, Cuesta1,Patiri1, 2006Basilakos_Shape}. The most extended
methodology along the literature, consists on performing statistics over
different ranges in the relative angle defined by the halo major axis and the
direction to larger structures defined via particular geometrical criteria. For
instance, \citet{2006Basilakos_Shape}, found that cluster sized halos are
aligned with their parent supercluster, in agreement with our results. The main
asset of our approach is that it allows a robust comparison between numerical
simulations and observational data, independently of any particular
characterisation of the surrounding structure. Given that the correlation
function can be easily extended from three dimensions to projected data, this
comparison arises naturally. 

Regarding the results in three dimensions, we have demonstrated that
cross-correlation functions with tracers in the directions parallel to the
minor and intermediate halo shape axes ($\xi_{\parallel c}$ and $\xi_{\parallel
b}$), show lower correlation amplitudes than the cross-correlation function
measured along the major axis ($\xi_{\parallel a}$) up to scales of $30$ Mpc
h$^{-1}$. We also found important differences between the parallel and
perpendicular cross-correlation functions (with respect to the major halo axis
$\hat a$). The detected correlation anisotropy signal was interpreted as a
decrement of the typical particle density in the surrounding matter
distribution in the plane perpendicular to the central halo major axis. We also
showed that the alignment signal is higher as the central halo sample mass
increases.

In order to confront simulations and observations, we have analysed the effect
of projections, group mass distribution and shot noise using the numerical
simulation. The alignment signal remains detectable, with an important
statistical significance level over a wide range of distances. 

The key result of the present study is the observational detection and
quantification of a large scale anisotropy around galaxy groups. In agreement
with full numerical simulations and mock catalogue results, our measurements of
the projected two point cross-correlation function for SDSS groups and
galaxies, exhibit an important degree of anisotropy. This signal is observed as
a difference in the correlation amplitude estimated in the directions parallel
and perpendicular to the group projected major axis. As expected, the amplitude
of this feature increases with the average mass of the groups. The observed
anisotropy is detected for group masses greater than $\simeq 6\times 10^{13}
{\rm M}_{\sun}$, and its statistical significance increases with the group
mass.

We average the alignment effect over ranges of scales corresponding to the $1-$
and $2-$halo terms in the cross-correlation function.  The $1-$halo term
results show similar anisotropies in the numerical simulation and the SDSS, and
are consistent with the known increase of asphericity as the group mass
increases \citep{PaperShapes}, a trend that is clearly detected in the
numerical simulation and only mildly present in the observational sample of
groups; this simply indicates that direct measurements of group shape (e.g.
$b/a$) are more suited to detect this trend.

The $2-$halo regime shows an excellent agreement between the mock model and
observations. We find that the alignment of the shape of groups with the
large-scale structure increases with the halo mass reflecting the fact that
higher mass groups have had, on average, more recent merger activity and
therefore, younger dynamical ages. This result is independent of the number of
group members used, indicating its lack of sensitivity to the shot-noise effect
that biases group shape estimates (i.e. it affects the shape more strongly than
the direction of the group major axis). The excellent quantitative agreement
between observations and the $\Lambda$CDM model results provides new direct
evidence of the adequacy of the gravitational instability to describe the
large-scale structure formation of our Universe.

\section*{Acknowledgements}

DP, MS and MM acknowledge support from CONICET and SECyT, Universidad Nacional
de C\'ordoba. DP acknowledges receipt of a postdoctoral fellowship from
CONICET. NP acknowledges support from Fondecyt Reg. 1071006, FONDAP CFA
15010003, and BASAL CATA PFB-06. 


\end{document}